\documentclass[aps,showpacs,amsmath,amssymb,floatfix,pre,twocolumn,superscriptaddress]{revtex4}
\usepackage{graphicx,amsmath}

\begin{document} 

\title {Symmetry breaking for ratchet transport in presence of interactions
and magnetic field}

\author{L. Ermann}
\affiliation{\mbox{Departamento de F\'\i sica Te\'orica, GIyA, 
Comisi\'on Nacional de Energ\'ia At\'omica, Buenos Aires, Argentina}}
\affiliation{\mbox{Laboratoire de Physique Th\'eorique du CNRS, IRSAMC, 
Universit\'e de Toulouse, UPS, F-31062 Toulouse, France}}

\author{A.D. Chepelianskii}
\affiliation{\mbox{Cavendish Laboratory, Department of Physics, 
University of Cambridge, CB3 0HE, United Kingdom}}

\author{D.L. Shepelyansky}
\affiliation{\mbox{Laboratoire de Physique Th\'eorique du CNRS, IRSAMC, 
Universit\'e de Toulouse, UPS, F-31062 Toulouse, France}}

\date{November 19, 2012}


\begin{abstract}
We study the microwave induced ratchet transport  
 of two-dimensional electrons 
on an oriented semidisk Galton board.
The magnetic field symmetries 
of ratchet transport are analyzed 
in presence of electron-electron interactions.
Our results show that a magnetic field asymmetric ratchet current 
can appear due to two contributions, a Hall drift of the rectified 
current that depends only weakly on electron-electron interactions 
and a breaking of the time reversal symmetry due to the combined 
effects of interactions and magnetic field. In the latter case, 
the asymmetry between positive and negative magnetic fields vanishes 
in the weak interaction limit.
We also discuss the recent experimental results 
on ratchet transport in asymmetric  nanostructures. 
\end{abstract}

\maketitle

\section{I. Introduction}
The appearance of a directed flow induced by a zero-mean monochromatic force
is a generic nonequilibrium phenomenon known
as the ratchet effect (see e.g. \cite{prost,hanggiphystod,reimann}).  
The ratchet transport at nanoscale attracts a significant interest in last years
(see \cite{hanggirmp} and Refs. therein).
The electron ratchet currents induced by a monochromatic {\it ac-}driving
have been experimentally observed in asymmetric mesoscopic structures
at high \cite{lorke}, very low \cite{linke} and GHz \cite{song} frequencies.
In the later experiments the ratchet effect was visible even at room temperatures
but these experiments were lacking a detailed analysis of magnetic field 
resonance effects, typical of two-dimensional electron gas (2DEG) in a magnetic field
\cite{weiss,kvon1} and a dependence study of ratchet current on microwave polarization.
We note that the presence of resonances in a resistivity 
dependence on a magnetic field and polarization dependence of ratchet transport
ensure that a voltage induced by a microwave radiation
appears due to an asymmetry of lattice structure and not due to
other asymmetries potentially 
present in an experimental devices.

The theoretical studies of 2DEG deterministic 
ratchet transport on a semidisk Galton board
has been started in \cite{alik2005} and further extended  in
\cite{cristadoro,alik2006},\cite{entin2006,entin2007}. The theoretical studies
show that the ratchet effect  exists not only for 2DEG but also for electrons
in graphene plane with oriented semidisk lattice \cite{ermann2011}.
The numerical simulations and analytical theory developed in 
\cite{alik2005,cristadoro},\cite{alik2006,entin2006},\cite{entin2007,ermann2011}
have been done for noninteracting electrons. 
They established that a directed
chaotic deterministic ratchet transport emerges in such asymmetric nanostructures
due to microwave radiation. The direction of ratchet current
can be controlled by the radiation polarization. 
The theoretical studies show that the
ratchet effect  also exists  in a 
presence of moderate and strong interactions between
electrons \cite{entin2008}. 

The theoretical works initiated the detailed experimental studies of the chaotic 
deterministic ratchet 2DEG transport on the semidisk Galton board
of antidots 
performed by Grenoble group \cite{portal1}. These experiments
clearly demonstrated the existence of  ratchet transport
in a high mobility 2DEG based on  AlGaAs/GaAs heterojunstions
with semidisk array. The polarization dependence of ratchet
current is found to be in a qualitative agreement with 
the theory dependence for noninteracting electrons.
It is also experimentally shown \cite{portal1} that the ratchet is absent in 
arrays with circular antidots ensuring that
the effect is produced by semidisks and not by always present 
device asymmetries. More recently the Grenoble group
performed ratchet experiments with 2DEG in Si/SiGe heterostructures
\cite{portal2,portal3}
where the interaction effects between electrons are expected to play
more important role \cite{entin2007}. The characteristic feature of these experiments
is the dependence of ratchet transport on a magnetic field
in presence of interactions. Thus it is important to understand the
properties of ratchet of interacting particles in 
a magnetic field which creates a symmetry breaking in space and time.
We note that previously the theoretical investigations were done
only for noninteracting electrons in a magnetic field 
\cite{alik2006,entin2007}
or for interacting electrons without magnetic field \cite{entin2008}. Thus
in this work we perform a more general study analyzing 
the properties of  
ratchet transport in presence of interactions and magnetic field.

Indeed, an applied magnetic field induces a chiral movement of charge in a conducting sample.
This chirality may be revealed in optical measurements such as the Faraday effect
\cite{faraday1,faraday2}. An emergence of a static magnetization due to 
an {\it ac-}electric field is known as the inverse Faraday effect (see e.g. \cite{polianski1}).
Hence, one can expect that transport properties of a chiral structure will strongly 
depend on the sign of magnetic field.
However, this argument, based on spatial symmetries, should also take into account 
that the equilibrium transport measurements are also constrained by
the time reversal symmetry that implies Onsager-Casimir reciprocity relations 
\cite{onsager,casimir}. 
Due to these relations a two terminal 
conductance is always symmetric with magnetic field masking the 
chirality. 
However, since the time reversal symmetry is valid only for equilibrium samples,
it can be destroyed for measurements in a nonlinear transport regime.
The appearance of magnetic field asymmetry in a nonlinear two terminal transport, 
has attracted a significant  theoretical and experimental attention 
in the mesoscopic physics community.
Indeed, a microscopic disorder potential has no symmetry,
and the absence of self-averaging in coherent samples makes such investigations possible.
Surprisingly, even if all symmetries are broken in the out of equilibrium regime 
theoretical calculations predict that interactions are required to observe a 
magnetic field antisymmetric component in nonlinear conductance. 
While measurements on coherent samples seem to support these predictions,
measurements on samples with artificial asymmetric structures 
may produce asymmetric (or even antisymmetric) nonlinear transport 
even in regimes where interactions do not seem to play a role (for e.g. high density)
\cite{portal1,lps}. Thus, the investigations of ratchet transport in presence of interactions and 
magnetic field will allow us to analyze the Onsager-Casimir reciprocity relations
in a new frame. 

In order to gain a deeper insight on the symmetry properties of nonlinear 
transport, we investigate numerically an interacting two-dimensional electron gas (2DEG) 
with a periodic array of asymmetric (semidisk) antidots oriented in a preferential direction
using the square lattice of oriented semidisks discussed in \cite{ermann2011,entin2008}.
In this model a semidisk of radius $r_d$ is placed in a square of size $R$
and then this square covers periodically the whole $(x,y)$ plane (see inset of Fig.~\ref{fig1}). 
We consider  $\;\;\;\;$ a homogeneous monochromatic  $\;\;\;\;$ linearly polarized $\;\;\;\;$ 
electric field 
${\bf E} = E_o \cos \omega t (\cos \theta, \sin \theta)$. 
This field creates 
a rectified current flow due to the asymmetric structure of the antidot 
superlattice.  The direction of the flow can be controlled by the polarization of the 
microwave field and by a magnetic field perpendicular to the 2DEG plane. 

The quantitative description of the ratchet current can be obtained 
on the basis of kinetic theory \cite{entin2007}.
However this theory rapidly breaks down in presence of a magnetic field since it does 
not capture the particle dynamics when cyclotron radius becomes of the order of the 
antidot radius.
Moreover it was shown recently \cite{entin2008}
that interactions could strongly modify the rectified current,
hence this system allows to investigate in detail the role of interactions on the magnetic 
field symmetry properties of the nonlinear response. 
In fact, we show that the antisymmetric component 
of the rectified current can be splitted in two terms.  
One term is weakly dependent on 
interactions and can be interpreted as 
a Hall drift of the rectified current in the preferential 
direction fixed by the semidisk superlattice.  
The second term vanishes in absence of 
interactions as predicted by theories \cite{Sanchez},\cite{Spivak},\cite{Polianski2}. 
Hence, depending on the measurement 
geometry, the antisymmetric component of the rectified current may vanish or not 
in absence of interactions. 

\section{II. Model description and results}

\begin{figure}[ht]
\begin{center}
\includegraphics[width=0.45\textwidth]{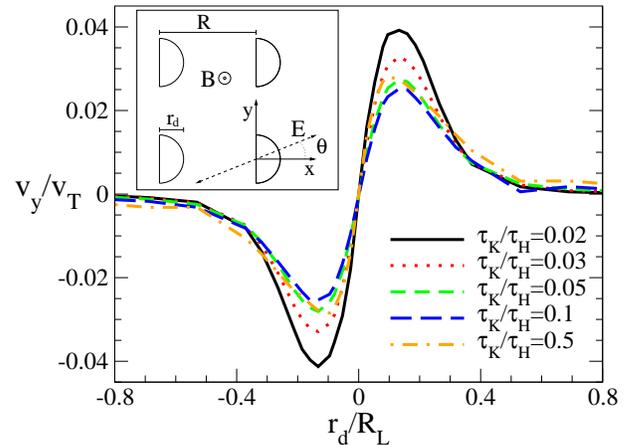}
\end{center}
\vglue 0.2cm
\caption{Dimensionless average ratchet velocity $v_y/v_T$ in direction $y$ as a function
of a magnetic field $B$ given by dimensionless parameter $r_d/R_L \propto B$. 
The electric field is linearly 
polarized in $y$ axis ($\theta=\pi/2$) with amplitude and 
frequency $E_0=0.1$ and $\omega=0.1$, and
the temperature of Hoover thermostat is $T=1$ with $\tau_H=10$.
The interactions between electrons goes from strong interaction regime ($\tau_K/\tau_H=0.02$) 
to weak interactions ($\tau_K/\tau_H=0.5$) as it is shown on legend.
The inset panel shows the directions of electric and magnetic fields
in a general case, and the
 distribution of semidisk periodic array with an orientation direction in $x$ axis. 
The geometric ratio is fixed at $R/r_d=4$ as it is shown on the inset.
\label{fig1}}
\end{figure}

In this work we consider a 2DEG with elastic semidisk scatterers of radius $r_d$, 
oriented in direction $\mathbf{e}_x=\mathbf{x}$, 
and placed in a periodic square lattice of size $R\times R$ (see inset of Fig~\ref{fig1}).
The electron motion is affected by an electric microwave field
 $\mathbf{E}\cos{\omega t}=E_0\left(\cos{\theta},\sin{\theta},0\right)\cos{\omega t}$ linear
polarized at angle $\theta$ to $\mathbf{e}_x$, and a transverse, 
uniform and constant magnetic 
field $\mathbf{B}\propto \mathbf{e}_z/R_L$ (inversely proportional to the Larmor radius $R_L$).
The system also interacts with a Nos\'e--Hoover thermostat  
which equilibrates the ensemble of particles 
to the Boltzmann distribution with temperature $T=mv_T^2/2$ in a characteristic 
time $\tau_H$ (see e.g. \cite{nosehoover}).
The electron interactions are treated in the frame of the mesoscopic multi-particle
collision model proposed by Kapral (see e.g, \cite{kapral}). 
The method consists in dividing the coordinate space of 
each square of size $R\times R$ with $N$ particles in $N_{cel}$ collision cells. 
Inside each cell the collisions are mimic with a rotation of all particle velocities 
on a random angle in the moving center-of-mass frame, preserving 
the total momentum and energy of the system. These rotations 
are done with a repetition period given by a characteristic time $\tau_K$.
In this way large and small values of $\tau_K$ correspond
to weak and strong interactions respectively. 
In this work we follow our previous studies of interactions effects on ratchet transport
\cite{entin2008} where the interactions were treated in the frame of Kapral approach. 

We fix the geometric ratio $R/r_d=4$ in order to work at low 
antidot density and to 
avoid geometrical particularities which may exist for $R/r_d\simeq1$.
The number of particles is fixed at  $N=10^4$, and for numerical simulations 
we choose dimensionless parameters 
$T=1$, $r_d=1$, $\tau_H=10$, and electron charge and mass $e=m_e=1$. 
The Kapral grid for interactions is fixed to have $N_{cel}=100\times100$
cells in the whole space region $R^2$. 
Therefore the only parameter controlling
the interaction strength is the inverse Kapral time $1/\tau_K$.
This choice of parameters is similar to those used in \cite{entin2008}.
We remind that $\tau_H$ determines the relaxation time to the equilibrium.
\begin{figure}[ht]
\begin{center}
\includegraphics[width=0.45\textwidth]{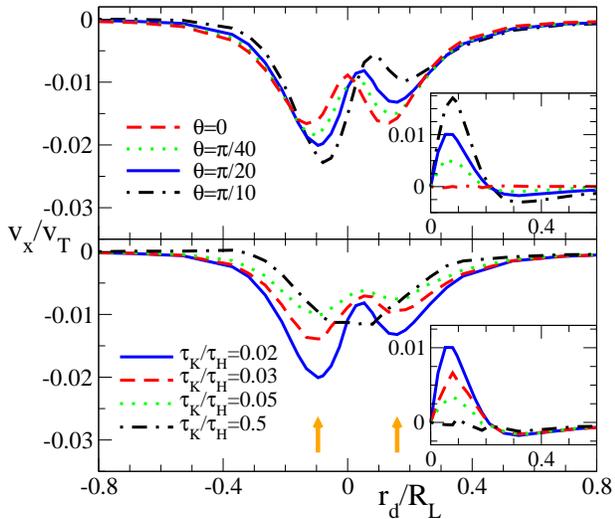}
\end{center}
\vglue 0.2cm
\caption{Dimensionless average ratchet velocity 
in direction $x$ ($v_x/v_T$) as a function of magnetic field.
{\it Top panel} shows the strong interactions regime (with $\tau_K/\tau_H=0.02$) 
for the rest of the parameters 
specified in caption of Fig~\ref{fig1}. 
The electrical field is polarized near $x$ axis, where dashed red curve represents the 
symmetrical case of $\theta=0$. The small deviation angles are shown at
$\theta=\pi/40$ by dotted green 
curve, $\theta=\pi/20$ by solid blue curve, and $\theta=\pi/10$ by dot-dashed black curve.
{\it Bottom panel} shows the case of fixed polarization angle $\theta=\pi/20$, 
and different strength of interactions between particles.
The interaction times are $\tau_K/\tau_H=0.02$ (solid blue curve), 
$\tau_K/\tau_H=0.03$ (dashed red curve), 
$\tau_K/\tau_H=0.1$ (dotted green curve) and $\tau_K/\tau_H=0.5$ (dot-dashed black curve).
{\it Inserts} in both panels show the dimensionless 
asymmetry $[v_x(r_d/R_L)-v_x(-r_d/R_L)]/v_T$
as a function of $r_d/R_L$ for the same parameters marked by color.
Orange arrows show the values of magnetic field analyzed in Fig. \ref{fig3}.
\label{fig2}}
\end{figure}

The steady-state net current of the system is proportional to the average particle velocities.
This current can be described as a vector $\mathbf{j}$ 
splitted in two components of different magnetic field symmetry: $\mathbf{j}=\mathbf{j}_s+\mathbf{j}_a$ 
where $\mathbf{j}_s$ are symmetric and $\mathbf{j}_a$ antisymmetric components. 
These two vectors depend quadratically on the alternating electric field amplitude $\mathbf{E}$.
and can also depend on $\mathbf{e}_x$ and $\mathbf{B}$, therefore their most general expression reads: 
\begin{eqnarray}
\label{eqs1}
 \mathbf{j}_s&=&f_1(B)\left(\mathbf{e}_x\cdot\mathbf{E}\right)\mathbf{E}+f_2(B)
\mathbf{E}^2\mathbf{e}_x\\
\nonumber
 \mathbf{j}_a&=&g_1(B)\left(\mathbf{e}_x\cdot\mathbf{E}\right)
\left(\mathbf{B}\wedge\mathbf{E}\right)+
g_2(B)\mathbf{E}^2\left(\mathbf{B}\wedge\mathbf{e}_x\right)
\end{eqnarray}
Here $f_i(B)$, $g_i(B)$ ($i=1,2$) are functions of the modulus of the magnetic 
field $B=|\mathbf{B}|$ depending implicitly on the rest of model parameters. 
Specifically, we focus on the dependence of the antisymmetric 
component of the current on the 
strength of interactions between electrons. 

We note that both terms of second line of
Eq.~\ref{eqs1} can be decoupled for different values of $\theta$.
For electric and magnetic fields used in this model, 
the dimensionless velocity can be written as
\begin{eqnarray}
\label{eqs2}
 \frac{v_x(\theta)}{v_T}&=&\frac{E_0^2}{\sqrt{2T}}\left[\tilde{f}_1\cos^2{\theta}+\tilde{f}_2-
\frac{\tilde{g}_1\sin{2\theta}}{2}\right]\\
\nonumber
 \frac{v_y(\theta)}{v_T}&=&\frac{E_0^2}{\sqrt{2T}}\left[\frac{\tilde{f}_1\sin{2\theta}}{2}+
\tilde{g}_1\cos^2{\theta}+\tilde{g}_2\right]
\end{eqnarray}
where $\tilde{f}_i$ and $\tilde{g}_i$ ($i=1,2$) are symmetric 
and antisymmetric functions of the 
magnetic field in $\mathbf{z}$ direction (or related Larmor radius $R_L=v_T/B$).
These functions are
proportional to the functions in 
Eqs.~\ref{eqs1} $\tilde{f}_i\propto f_i(B)$ and $\tilde{g}_i\propto g_i(B)/R_L$ 
(where we take $R_L>0$  and $R_L<0$ for magnetic field in $\mathbf{z}$ and $-\mathbf{z}$ 
directions respectively).
Following Eqs.~\ref{eqs2} we can note that for $\theta=0$ and $\theta=\pi/2$ velocities 
in $x$ direction are symmetric while the $y$-components are antisymmetric.

In the case of polarized electric field in $y$ direction ($\theta=\pi/2$), $v_y$ 
can be asymmetric in 
the magnetic field only due to the second term of second Eq.~\ref{eqs2}
noted as $\tilde{g}_2$.
This case is illustrated in Fig.~\ref{fig1} for 
the dimensionless quantities $v_y/v_T$ plotted versus $r_d/R_L$ 
at different values of Kapral time $\tau_K$ and 
the rest of parameter specified in the caption.
The data of Fig.~\ref{fig1} show that the dependence of ratchet velocity $v_y/v_T$
is an asymmetric function of $r_d/R_L \propto B$ with maximum of $|v_y|$ at
$r_d/R_L \approx \pm 0.15$. The amplitude of this maximum increases 
by a factor 2 with 
the increase of interactions (decrease of $\tau_K$).

Following the first line of Eq.~\ref{eqs2}
the asymmetry given by $\tilde{g}_1$ can be analyzed via parameter dependence of
$v_x/v_T$  at small angles of linear polarization 
of electric field directed both along $\mathbf{x}$ and $\mathbf{y}$.
The emergence of $\tilde{g}_1$ from a symmetric behavior in the magnetic field is shown in 
top panel of Fig.~\ref{fig2}, for $v_x/v_T$ with small values 
of $\theta$ in the strong interactions regime ($\tau_K/\tau_H=0.02$).
In the case of $\theta=\pi/20$, bottom panel of Fig.~\ref{fig2} shows the behavior of 
asymmetry in $v_x/v_T$ for different interaction times, going from $\tau_K/\tau_H=0.02$ 
to $\tau_K/\tau_H=0.5$. 
The striking feature of bottom panel of Fig.2 
is that the asymmetry in a magnetic field is rather strong at strong interactions
while in the limit of weak interactions 
the asymmetry completely disappears and we recover the symmetric curve
as a function of a magnetic field.

\begin{figure}[ht]
\begin{center}
\includegraphics[width=0.235\textwidth]{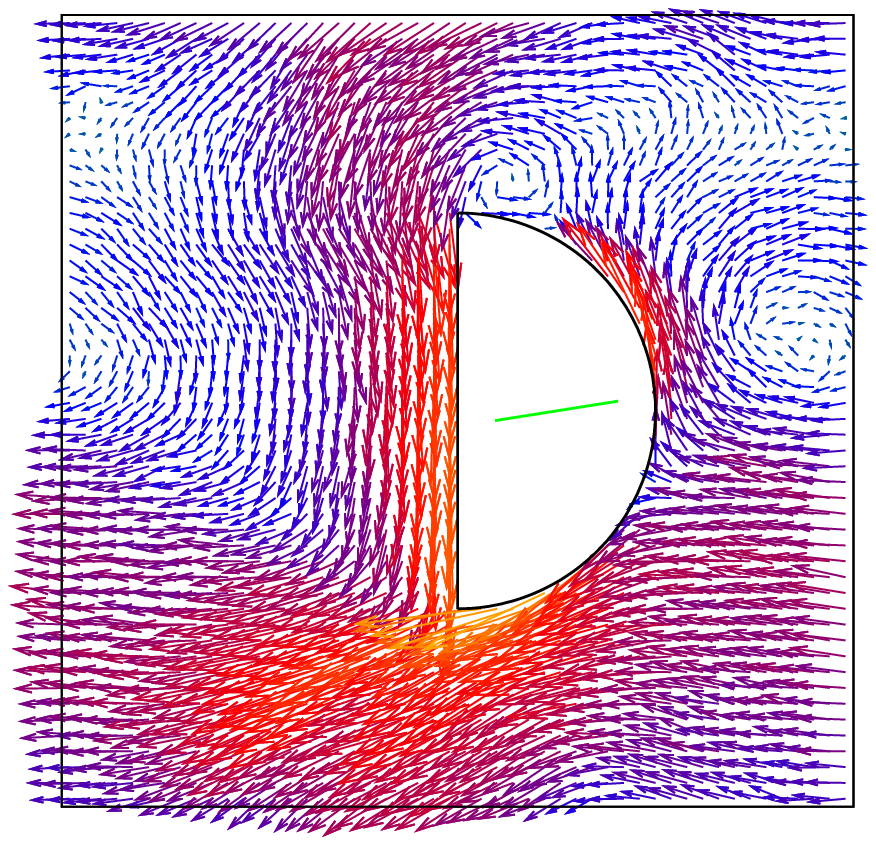}
\includegraphics[width=0.235\textwidth]{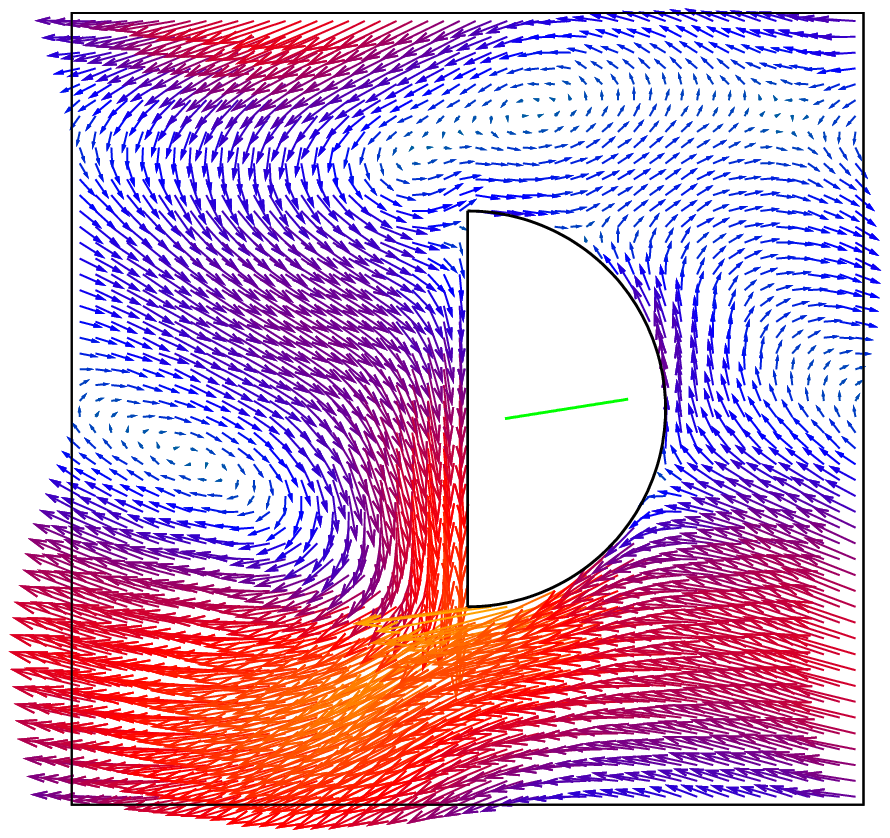}\\
\includegraphics[width=0.235\textwidth]{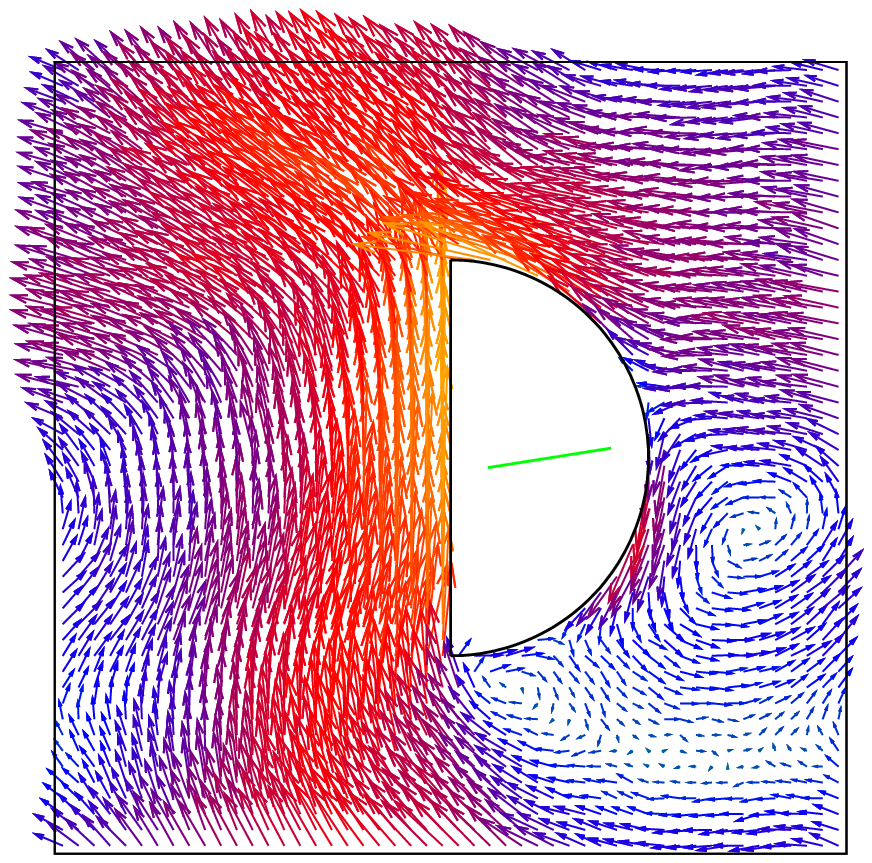}
\includegraphics[width=0.235\textwidth]{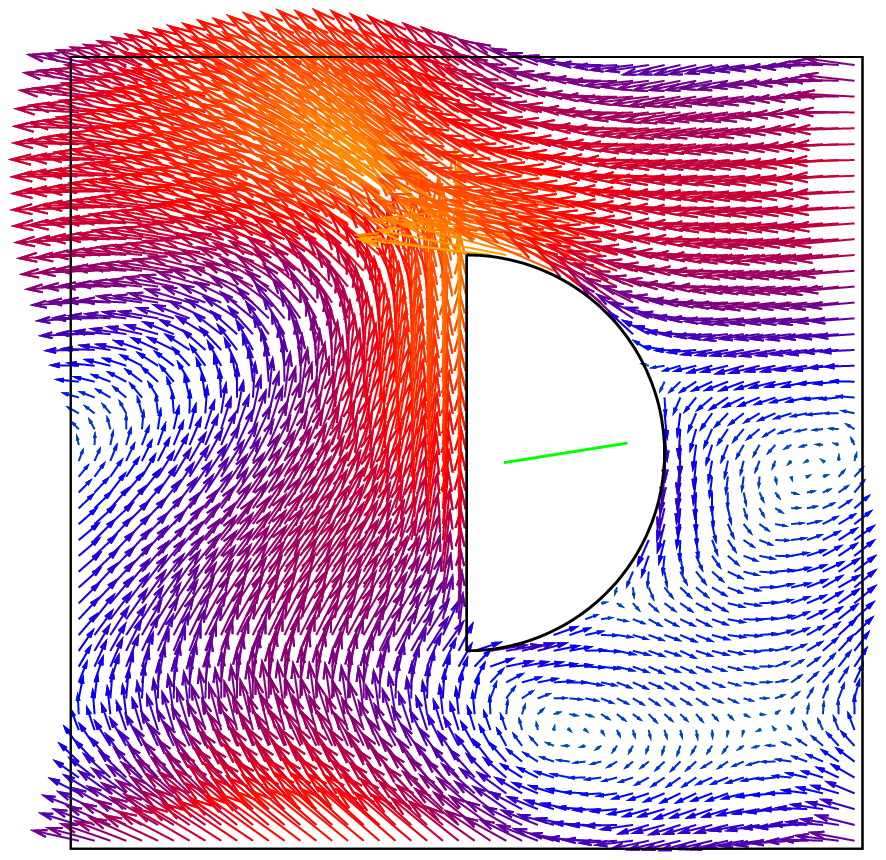}
\end{center}
\vglue 0.2cm
\caption{Map of local averaged velocities in $(x,y)$ with $x,y\in[-3,3)$ and 
linear polarized electrical field $\theta=\pi/20$ (shown by
green line inside semidisk scatterer);
other parameters are the same as  in Fig.~\ref{fig1}.
The left panels show the cases of weak interactions between particles 
at $\tau_K/\tau_H=0.5$, 
while the right panels show strong interactions regime at $\tau_K/\tau_H=0.02$. 
The values of magnetic field are given by $r_d/R_L=0.1591$ on top panels, and  $r_d/R_L=-0.0955$ 
on bottom panels (pointed with orange arrows in Fig. \ref{fig2}).
The velocities are shown by arrows which size is proportional to velocity amplitudes, which is 
also indicated by color [from yellow (gray) for large to blue (black) for small amplitudes].
\label{fig3}}
\end{figure}

A deeper analysis of the symmetry properties can be done by
considering the flux of velocities in coordinate space.
It is known that for interacting particles the average velocity behavior can be 
rather complex with an emergence of  some vortexes  \cite{entin2008}. 
In Fig.~\ref{fig3} we present the velocity flux 
for weak and strong interactions regimes with positive and negative magnetic fields.
The analyzed values of Larmor radius correspond to the relative minima of $v_x/v_T$ marked 
with arrows in Fig.~\ref{fig2} at negative and positive values of $r_d/R_L$. 
Fig.~\ref{fig3} shows the flow structure at $r_d/R_L\simeq 0.16$
on top panels and at $r_d/R_L\simeq -0.1$ on bottom panels.
The polarization angle in four panels is fixed at $\theta=\pi/20$, and therefore the reflection 
symmetry $y\rightarrow-y$ is not preserved.
The data show that the vortex asymmetric structure
 is more pronounced in the case of strong interactions
on right panels.

\begin{figure}[ht]
\begin{center}  
\includegraphics[width=.475\textwidth]{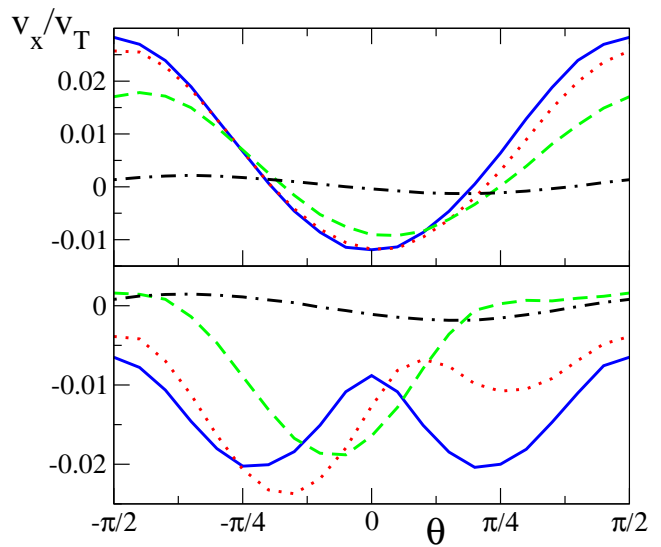}
\caption{(color online) Polarization dependence of 
$x$-component of the averaged ratchet velocity $v_x/v_T$, 
for weak interactions between particles in top panel ($\tau_K/\tau_H=0.5$), and 
for strong interactions in bottom panel ($\tau_K/\tau_H=0.02$). 
Four different values of magnetic fields are shown in both panels: zero magnetic field 
($r_d/R_L=\infty$, solid blue curve), $r_d/R_L=0.053$ by dotted red curve, 
$r_d/R_L=0.133$ by dashed green curve and $r_d/R_L=0.53$ by dot-dashed black curve.}
\label{fig4}
\end{center}
\end{figure}

The polarization dependence of ratchet $\; \;$ current in  $\; \;$ $x-$direction is 
analyzed in Fig.~\ref{fig4}. We see that even for various magnetic fields
the dependence on polarization angle  $\theta$ is essentially symmetric
at weak interactions (top panel)
while at strong interactions (bottom panel) we have a strongly asymmetric behavior
at moderate magnetic fields ($r_d/R_L=0.053, 0.133$). At a relatively large magnetic field
($r_d/R_L=0.53$) the amplitude of ratchet current becomes rather small
since the Larmor radius becomes smaller than the distance between
semidisks. Thus the data of Fig.~\ref{fig4} also show that the asymmetry of 
ratchet transport appears only in a presence of interactions. 

\section{III. Discussion}

In this work we study the symmetry properties of the ratchet transport
on an oriented semidisk antidot superlattice.
We show that while the ratchet flow on the oriented 
semidisk superlattice (along $x-$axis)
does not depend on the sign of  magnetic field for a noninteracting 2DEG, 
interactions can give rise to an antisymmetric component of the flow in the semidisk direction
as a function of a magnetic field.
This result is consistent with the case of mesoscopic samples where deviations from Onsager-Casimir 
reciprocity relations were shown to occur only in presence 
of electron-electron interactions \cite{Sanchez},\cite{Spivak},\cite{Polianski2}.
On the contrary, the flow perpendicular to the semidisk direction ($y-$axis) is 
asymmetric even when interactions are absent. We argue that the origin of this component of the flow
is a Hall drift of the rectified current. This contribution is absent in disordered mesoscopic 
samples because the translational symmetry is broken by a disorder potential 
where no preferential direction is present for the flow. 

The comparison with  \cite{portal1},\cite{portal2},\cite{portal3}
highlights various aspects of these skillful experiments.
At weak electron-electron interactions typical of 
AlGaAs/GaAs hererojunctions \cite{portal1}
there is a qualitative agreement between 
the theory and experiment on polarization dependence
and approximately symmetric current response on sign change of a magnetic field,
even if the absolute quantitative  values differ from theoretical predictions (see \cite{entin2008}
for a more detailed discussion). For experiments with Si/SiGe heterostructures 
\cite{portal2,portal3} the effects of interactions are stronger
and asymmetric response in a magnetic field is clearly observed
in experiments (see e.g. Fig.3 in \cite{portal2}). In this case the
change of polarization from $\theta=0$ to $\theta=\pi/2$
does not change  the sign of photovoltage that also appears
in theoretical models at strong interactions as discussed in \cite{entin2008}.
Also the ratchet effect is clearly suppressed in experiment and theory
at large magnetic fields.
However, at the same time there are  significant differences between
experiments \cite{portal2,portal3} and theoretical results 
presented here. Indeed, for $\theta=\pi/2$ the theory 
predicts a change of the photovoltage sign upon inversion of 
the magnetic field (see Fig.~\ref{fig1} above)
while there is no such sign change in experiments
(see e.g. Fig.3 in \cite{portal2} and in \cite{portal3}).
It is possible that finite sample size leads to a certain charge accumulation
in the experimental setup (see indications for that in Fig.4 in \cite{portal3})
that may explain the difference with the theory
where the analysis is done for an infinite lattice size. 
We should also note that the present studies were done for a square lattice of semidisks
while in the experiments \cite{portal1,portal2,portal3}
were performed on an hexagonal lattice of semidisks.
However, in both cases the density of semidisks
is relatively low (since $R/r_d=4$ here and
$R/r_d \approx 5$ in \cite{portal2,portal3}).
Thus, the ratchet transport is created mainly by scattering on a one semidisk
(see discussion in \cite{entin2007}) and the difference between 
square and hexagonal lattices is not expected to be important.

The discussed caparison between the present theoretical studies and 
the most advanced experiments reported in \cite{portal1},\cite{portal2},\cite{portal3} 
shows that the further experimental and theoretical research 
of the electron ratchet transport in asymmetric nanostructures 
with dynamical chaos represents a significant 
fundamental scientific interest.

We thank M.L. Polianski for fruitful discussions on out of equilibrium transport in mesoscopic systems.
This work was supported in part by the French PNANO ANR project NANOTERRA, one of us (A.D.C.) 
acknowledges support from St Catharine's College.

\end{document}